\documentstyle[12pt,epsfig]{article} 

\begin{document}  
\vspace*{-2cm}  
\renewcommand{\thefootnote}{\fnsymbol{footnote}}  
\begin{flushright}  
hep-ph/0006077\\
PSI-PR 00 10\\  
June 2000\\  
\end{flushright}  
\vskip 45pt  
\begin{center}  
{\Large \bf Mass gap effects and higher order 
electroweak Sudakov logarithms}\\
\vspace{1.2cm} 
{\bf  
Michael Melles\footnote{Michael.Melles@psi.ch}   
}\\  

\begin{center}
Paul Scherrer Institute (PSI), CH-5232 Villigen, Switzerland. 
\end{center}

\vspace{20pt}  
\begin{abstract}
The infrared structure of spontaneously broken gauge theories is 
phenomenologically very important and theoretically a challenging problem.
Various attempts have been made to calculate the higher order behavior
of large double-logarithmic (DL) corrections originating from the exchange of
electroweak gauge bosons resulting in contradictory claims. 
We present results from two loop electroweak 
corrections for the process $g \longrightarrow f_{\rm
R} {\overline f}_{\rm L}$ to DL accuracy. This process is ideally suited
as a theoretical model reaction to study the effect of the mass gap of the
neutral electroweak gauge bosons at the two loop level. Contrary to recent
claims in the literature, we find that the calculation performed with the
physical Standard Model fields is in perfect agreement with the results
from the infrared evolution equation method. In particular, we can confirm
the exponentiation of the electroweak Sudakov logarithms through two loops.
\end{abstract}
\end{center}  
\vskip12pt

\setcounter{footnote}{0}  
\renewcommand{\thefootnote}{\arabic{footnote}}  
  
\vfill  
\clearpage  
\setcounter{page}{1}  
\pagestyle{plain} 
 
\section{Introduction} 

Recently there have been various attempts to calculate higher order
corrections in the electroweak theory with differing outcomes 
\cite{flmm,cc,kp,kps,m}.
At future collider experiments in the TeV regime, large double logarithmic
(DL) corrections originating from the exchange of the Standard Model (SM)
gauge bosons need to be taken into account for precision measurements
designed to disentangle new physics effects. Even two loop effects are
expected to change typical cross sections by a few percent. The reason for
these large effects is simply that cross sections in the electroweak theory
depend on the infrared cutoff, the gauge boson mass $M$, and don't cancel
out in semi-inclusive, or even in fully inclusive cross sections \cite{ccc}.

Thus it is of considerable importance and of theoretical interest to study
the higher order behavior of these Sudakov logarithms in general scattering
amplitudes. In Ref. \cite{flmm} a general method of resumming these
large DL corrections has been presented based on the gauge invariant
infrared evolution equation method \cite{kl}. In Ref. \cite{m} the approach was
extended to the subleading level and shown to agree with explicit one
loop calculations.
In both cases these corrections were found to exponentiate also in the
case of broken gauge theories. Recently, however, an explicit two loop
calculation \cite{bw} in the Coulomb gauge found non-exponentiating contributions
for external fermion lines originating from a mass gap 
between the photon and Z-boson.

It is the purpose of this paper to refute these claims and also to demonstrate
with a calculation in terms of the physical fields, that exponentiation
of electroweak Sudakov logarithms holds at the two loop level.
For this purpose we study the reaction $g \longrightarrow f_{\rm R}
\overline{f}_{\rm L}$, which is ideally suited to study the
mass gap effects from the exchange of the neutral electroweak gauge bosons. 

In the next section we briefly review the calculation of the one loop
DL corrections in the Feynman gauge. This gauge leads to a particularly
simply way of calculating DL corrections and only QED-like topologies have
to be considered in our example. The important difference to QED is of course
the fact that the two neutral gauge bosons have different masses.
These effects are then investigated at the two loop level using the 
physical SM fields.
We only consider virtual corrections with the understanding that for
physical cross sections also the real soft emission contributions have
to be included.
We then compare our result with the literature and summarize our findings
briefly.
\section{Two loop results} 

We begin by briefly recalling the Sudakov method \cite{s}
of extracting DL corrections
from Feynman diagrams\footnote{More details, including the effect of fermion
masses, are given in Ref. \cite{ms}.}. In the Feynman gauge, the propagator of the exchanged
boson must connect two external lines in order to generate the 
double pole structure needed to obtain DL corrections.
The large argument in the logarithms is $s=2p_1p_2$, where we denote with 
$p_1$ the four momentum of $f_{\rm R}$ and with $p_2$ the four momentum of 
${\overline f}_{\rm L}$. The Sukakov decomposition along these external
momenta is given by
\begin{equation}
k=vp_2 \; + u p_1 + k_\perp \label{eq:sudp}
\end{equation}
For the boson propagator we use the identity
\begin{equation}
\frac{i}{s u v - m^2-{\mbox{\boldmath $k$}^2_{\perp}}+i\varepsilon}= {\cal P}
\frac{i}{s u v - m^2-{\mbox{\boldmath $k$}^2_{\perp}}} + \pi \delta ( s u v - 
m^2-{\mbox{\boldmath $k$}^2_{
\perp}}) \label{eq:propid}
\end{equation}
writing it in form of the real and imaginary parts (the principle value is indicated
by ${\cal P}$).
The latter does not contribute to the DL asymptotics and at higher orders gives
subsubleading
contributions.
 Rewriting the measure as $d^4k=d^2k_\perp d^2k_\parallel$ with
 \begin{eqnarray}
 d^2k_\perp &=& |{\mbox{\boldmath $k$}_{\perp}}| d |{\mbox{\boldmath $k$}_{\perp
 }}| d \phi =
 \frac{1}{2} d {\mbox{\boldmath $k$}^2_{\perp}} d \phi = \pi d {\mbox{\boldmath
 $k$}^2_{\perp}}
 \\
 d^2k_\parallel &=& | \partial (k^0,k^x)/ \partial (u,v)| d u d v \approx \frac{
 s}{2} du dv
 \end{eqnarray}
 where we turn the coordinate system such that the $p_1,p_2$ plane corresponds to
$0,x$ and the
 $y,z$ coordinates to the $k_\perp$ direction so that it is purely spacelike.
 The last equation follows from $p_i^2=0$, i.e. $p_{i_x}^2\approx p_{i_0}^2$ and

 \begin{equation}
 (p_{1_0}p_{2_x}-p_{2_0}p_{1_x})^2 \approx (p_{1_0}p_{2_0}-p_{2_x}p_{1_x})^2=(p_1
 p_2)^2=s/2
 \end{equation}
\begin{figure}
\centering
\epsfig{file=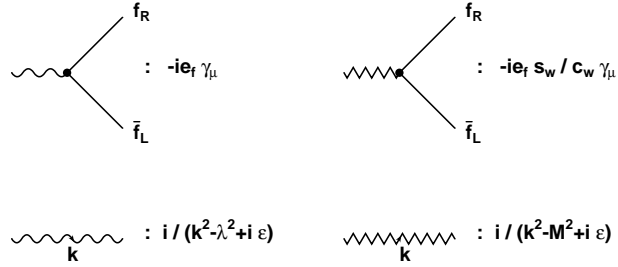,width=8cm}
\caption{The electroweak SM Feynman rules employed in this work.
For the propagators (modulo $g^{\mu \nu}$), we use the Feynman gauge.
The neutral Z-boson couples to right handed fermions through a rescaled
QED-like coupling. The Dirac-algebra is therefore identical to the Abelian
case.} \label{fig:ewfr}
\end{figure}
\begin{figure}
\centering
\epsfig{file=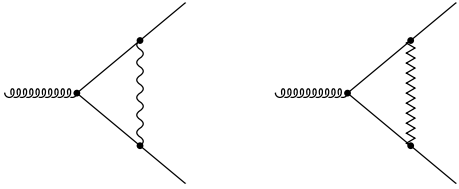,width=6cm}
\caption{The one loop electroweak SM Feynman diagrams leading to DL corrections
in the Feynman gauge
for $g \longrightarrow f_{\rm R} {\overline f}_{\rm L}$. 
Only the vertex corrections from the neutral Z-boson (zigzag-lines) 
and the photon propagators contribute. At higher orders only corrections to
these two diagrams need to be considered in the DL approximation. The 
photonic corrections are regulated by a fictitious mass terms $\lambda$.
In physical cross sections, the $\lambda$-dependence is canceled by the
effect of the emission of soft and collinear bremsstrahlung photons.}
\label{fig:ew1l}
\end{figure}
At the one loop level, the electroweak corrections are
depicted in Fig.~\ref{fig:ew1l}. We use the Feynman gauge and thus the
Feynman rules depicted in Fig.~\ref{fig:ewfr}. The fermion masses are
neglected for simplicity. They can, however, be added without changing the
nature of the higher order corrections.
For right handed fermions we only need to consider the neutral electroweak
gauge bosons, i.e. we are concerned with an $U(1)_{\rm R} \times
U(1)_{\rm Y}$ gauge theory which is spontaneously broken to yield
the Z-boson and photon fields. 
The DL-contribution of a particular Feynman diagram is thus given by
\begin{equation}
{\cal M}_k = {\cal M}_{\rm Born} \; {\cal F}_k \label{eq:formfac}
\end{equation}
where the ${\cal F}_k$ are given by integrals over the remaining 
Sudakov parameters
at the $n$-loop level:
\begin{equation}
\quad {\cal F}_k = \left( \frac{ \alpha_f
}{2 \pi}
\right)^n \prod^n_{i=1} \int^1_0 \int^1_0 \frac{d u_i}{u_i} \frac{d
v_i}{v_i} \Theta_k \label{eq:SudQED} 
\end{equation}
The $\Theta_k$ describe the regions of integration which lead to DL corrections. 
At one loop, the diagrams of Fig.~\ref{fig:ew1l} lead to
\begin{eqnarray}
{\cal M}^{{\rm R}^{(1)}}_{\rm DL} &=& {\cal M}^{\rm R}_{\rm Born} \left( 1 - 
\frac{\alpha_f}{2 \pi} \int^1_0 \frac{du}{u}
\int^1_0 \frac{dv}{v} \left[ \theta (s u v- \lambda^2)+ 
\frac{\rm s_w^2}{\rm c_w^2} \theta ( s u v - M^2) \right] \right) \nonumber \\
&=&{\cal M}^{\rm R}_{\rm Born} \left( 1
- \frac{\alpha_f}{4 \pi} \left[ \log^2 \frac{s}{\lambda^2}+
\frac{\rm s_w^2}{\rm c_w^2} \log^2 \frac{s}{M^2} \right] \right) \label{eq:ew1l}
\end{eqnarray}
which is the well known result from QED plus the same term with a rescaled
coupling (see Fig.~\ref{fig:ewfr}) and infrared cutoff. The restriction
to right handed fermions allows us to focus solely on the mass gap of
the neutral 
electroweak gauge bosons. The $W^\pm$ only couples to left handed doublets.
\begin{figure}
\centering
\epsfig{file=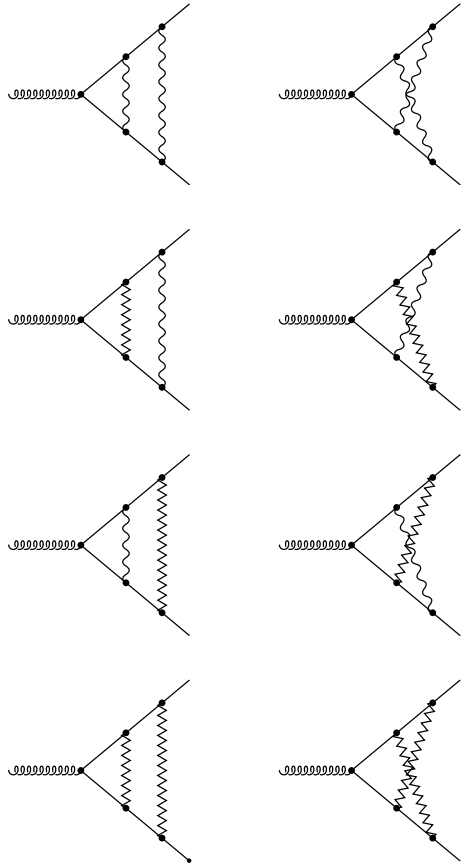,width=6cm}
\caption{The two loop electroweak SM Feynman diagrams leading to DL corrections
in the Feynman gauge
for $g \longrightarrow f_{\rm R} {\overline f}_{\rm L}$. 
The neutral Z-boson (zigzag-lines) and the photon propagators possess 
different on-shell regions due to the mass gap.} \label{fig:ewfd}
\end{figure}
At the two loop level we have to consider more diagrams than in the
QED case. The relevant Feynman graphs that give DL corrections in the
Feynman gauge are depicted in Fig.~\ref{fig:ewfd}. 
Only these corrections can yield four logarithms at the two loop level
in the Feynman gauge. Otherwise one cannot obtain the required pole terms
(as is well known in QED \cite{ms}).
They contain diagrams
where the exchanged gauge bosons enter with differing on-shell regions,
i.e. differing integration regions which give large DL corrections.
It is instructive to revisit
the case of pure QED corrections, since the topology of the graphs
yielding DL contributions in the Feynman gauge is unchanged. 
In QED at the two loop level, the scalar integrals corresponding
to the first row of Fig.~\ref{fig:ewfd} are given by:
\begin{eqnarray}
S^{\rm QED}_1 \!\!\!\!\!&\equiv& \!\!\!\!\! \int^1_0 \! \frac{du_1}{u_1} \!
\int^1_0 \! \frac{dv_1}{v_1} \!
\int^1_0 \! \frac{du_2}{u_2} \!
\int^1_0 \frac{dv_2}{v_2} \theta ( s u_1 v_1 - \lambda^2) \theta (s u_2 v_2
- \lambda^2) \theta ( u_1\!\!-\!\!u_2) \theta (v_1\!\!-\!\!v_2) \label{eq:qed1} \\
S^{\rm QED}_2 \!\!\!\!\!&\equiv&\!\!\!\!\! 
\int^1_0 \! \frac{du_1}{u_1} \! \int^1_0 \! \frac{dv_1}{v_1} \! \int^1_0 \!
\frac{du_2}{u_2} \!
\int^1_0 \! \frac{dv_2}{v_2} \theta ( s u_1 v_1 - \lambda^2) \theta (s u_2 v_2
- \lambda^2) \theta ( u_1\!\!-\!\!u_2) \theta (v_2\!\!-\!\!v_1) \label{eq:qed2}
\end{eqnarray}
Thus, in QED we find the familiar result
\begin{eqnarray}
S^{\rm QED}_1+S^{\rm QED}_2 &=&
\int^1_0 \! \frac{du_1}{u_1} \!
\int^1_0 \! \frac{dv_1}{v_1} \!
\int^1_0 \! \frac{du_2}{u_2} \!
\int^1_0 \frac{dv_2}{v_2} \theta ( s u_1 v_1 - \lambda^2) \theta (s u_2 v_2
- \lambda^2) \theta ( u_1-u_2) \nonumber \\
&=& \int^1_\frac{\lambda^2}{s} \frac{du_1}{u_1} \int^{1}_\frac{\lambda^2}{
s u_1} \frac{dv_1}{v_1} \frac{1}{2} \log^2 \frac{s u_1}{\lambda^2} 
\nonumber \\
&=& \frac{1}{2} \left( \frac{1}{2} \log^2 \frac{s}{\lambda^2} \right)^2
\label{eq:qedsud}
\end{eqnarray}
whichs yields the second term of the exponentiated one loop result in
Eq.~(\ref{eq:ew1l})
for ${\rm s_w} \longrightarrow 0$.
In the electroweak theory, we also need to consider the remaining diagrams
of Fig.~\ref{fig:ewfd}. The only differences occur because of the rescaled
coupling according to the Feynman rules in Fig.~\ref{fig:ewfr} and the
fact that the propagators have a different mass. Thus the second row
of Fig.~\ref{fig:ewfd} leads to
\begin{eqnarray}
S^{\rm M, \lambda}_1+S^{\rm M, \lambda}_2 &=& \frac{\rm s^2_w}{\rm c^2_w}
\int^1_0 \! \frac{du_1}{u_1} \!
\int^1_0 \! \frac{dv_1}{v_1} \!
\int^1_0 \! \frac{du_2}{u_2} \!
\int^1_0 \frac{dv_2}{v_2} \theta ( s u_1 v_1 - M^2) \theta (s u_2 v_2
- \lambda^2) \theta ( u_1-u_2) \nonumber \\
&=& \frac{\rm s^2_w}{\rm c^2_w}
\int^1_\frac{M^2}{s} \frac{du_1}{u_1} \int^{1}_\frac{M^2}{
s u_1} \frac{dv_1}{v_1} \frac{1}{2} \log^2 \frac{s u_1}{\lambda^2} 
\nonumber \\
&=& \frac{\rm s^2_w}{\rm c^2_w}
\left[ \frac{1}{8} \log^4 \frac{s}{\lambda^2} - \frac{1}{6} \log^3 
\frac{s}{\lambda^2} \log \frac{M^2}{\lambda^2} 
+ \frac{1}{24} \log^4 \frac{ M^2}{\lambda^2} \right]
\label{eq:2sud}
\end{eqnarray}
where we indicate the gauge boson masses of the two propagators in the scalar
functions. Analogously we find for the remaining two rows 
\begin{eqnarray}
S^{\rm \lambda, M}_1+S^{\rm \lambda, M}_2 &=& \frac{\rm s^2_w}{\rm c^2_w}
\int^1_0 \! \frac{du_1}{u_1} \!
\int^1_0 \! \frac{dv_1}{v_1} \!
\int^1_0 \! \frac{du_2}{u_2} \!
\int^1_0 \frac{dv_2}{v_2} \theta ( s u_1 v_1 - \lambda^2) \theta (s u_2 v_2
- M^2) \theta ( u_1-u_2) \nonumber \\
&=& \frac{\rm s^2_w}{\rm c^2_w}
\int^1_\frac{M^2}{s} \frac{du_2}{u_2} \int^{1}_\frac{M^2}{
s u_2} \frac{dv_2}{v_2} \frac{1}{2} \left(\log^2 \frac{s}{\lambda^2} -
\log^2 \frac{s u_2}{\lambda^2} \right) 
\nonumber \\
&=& \frac{\rm s^2_w}{\rm c^2_w}
\left[ \frac{1}{4} \log^2 
\frac{s}{M^2} \log \frac{ s}{\lambda^2} - \left( 
S^{\rm M, \lambda}_1+S^{\rm M, 
\lambda}_2 \right) \right]
\label{eq:3sud} \\
{\rm and} \;\;\;\;\;\;\;\;\;\;\;\;\;\;\; &&  \nonumber \\
S^{\rm M, M}_1+S^{\rm M, M}_2 &=& 
\frac{1}{2} \frac{\rm s_w^4}{\rm c_w^4} \left( \frac{1}{2} \log^2 \frac{s}{M^2} \right)^2
\label{eq:4sud}
\end{eqnarray}
Thus, we find for the full two loop electroweak DL-corrections  
\begin{eqnarray}
{\cal M}^{{\rm R}^{(2)}}_{\rm DL} &\equiv& {\cal M}^{\rm R}_{\rm Born}\left( 1 + 
\delta_{\rm R}^{(1)} + \delta_{\rm R}^{(2)} \right) \label{eq:2lres} \\
{\rm with} \;\;\;\;\;\;\;\;\;\; && \nonumber \\
\delta_{\rm R}^{(1)} &=& 
- \frac{\alpha_f}{4 \pi} \left( \log^2 \frac{s}{\lambda^2} + \frac{\rm s^2_w}{\rm c^2_w}
\log^2 \frac{s}{M^2} \right) \\ 
{\rm and} \;\;\;\;\;\;\;\;\;\; && \nonumber \\
\delta_{\rm R}^{(2)}&=& \left(\frac{\alpha_f^2}{2 \pi} \right)^2 \left[ 
S^{\rm QED}_1+S^{\rm QED}_2+S^{\rm M, \lambda}_1+S^{\rm M,\lambda}_2
+S^{\rm \lambda, M}_1+S^{\rm \lambda, M}_2 + S^{\rm M, M}_1+S^{\rm M, M}_2
\right] \nonumber \\
&=& \left(\frac{\alpha_f^2}{2 \pi} \right)^2 \left[
\frac{1}{8} \log^4 \frac{s}{\lambda^2} +
\frac{\rm s^2_w}{4 \rm c^2_w} \log^2 \frac{s}{M^2} \log^2 \frac{ s}{\lambda^2}
+\frac{\rm s^4_w}{8 \rm c^4_w} \log^4 \frac{s}{M^2} \right] \nonumber \\
&=& \frac{1}{2} \left[- \frac{\alpha_f}{4 \pi} \left( \log^2 \frac{s}{\lambda^2} + \frac{\rm s^2_w}{\rm c^2_w}
\log^2 \frac{s}{M^2} \right) \right]^2 \label{eq:ewres}
\end{eqnarray}
which is precisely the second term of the exponentiated one loop result
in the process of $g \longrightarrow f_{\rm R} {\overline f}_{\rm L}$. 
In the next section we review briefly the infrared evolution equation method
and compare Eq.~(\ref{eq:2lres}) with other results in the literature.

\section{Comparison with other results} 

In Ref. \cite{flmm} a general method for calculating DL corrections in the
electroweak theory was presented. It is based on calculating the DL corrections
first in the high energy regime $s \gg \mu^2 \ge M^2$, where $\mu^2$ is an
infrared cutoff on the ${\bf k}_\perp^2$ of the exchanged gauge bosons.
The corrections are then obtained by solving the infrared evolution equation,
based on a non-Abelian generalization of Gribov's bremsstrahlung theorem
\cite{vg}.
The solution in this regime is then taken as the initial condition for the
general case where $M^2 \leq \mu^2$. In this way the mass gap between 
the neutral electroweak gauge bosons is incorporated in a natural way by
considering the effective theories in each domain and requiring equality
at the weak scale $M$.
In Ref. \cite{m} an alternative derivation was presented based on the virtual
contributions to the Altarelli-Parisi splitting functions.
In this approach it is evident that all Sudakov logarithms, for fermions and
transversely polarized gauge bosons even on the subleading level, are
related to external lines and exponentiate. For longitudinally polarized
gauge bosons all DL corrections can be resummed by utilizing the Goldstone
boson equivalence theorem \cite{m}.

In order to compare the result of Refs. \cite{flmm,m} with Eq.~(\ref{eq:2lres}) we
specialize on the case with an external right handed fermion and left handed
anti-fermion pair. In terms of the quantum numbers of the unbroken theory,
we obtain from Refs. \cite{flmm,m}
\begin{eqnarray}
{\cal M}^{\rm R}_{\rm DL} &=& {\cal M}^{\rm R}_{\rm Born} 
\exp \left( - \frac{1}{2} \sum^2_{k=1} \left[ \frac{g^2}{(4 \pi)^2} T_k (T_k+1)
+\frac{{g^\prime}^2}{(4 \pi)^2} \frac{Y_k^2}{4} \right] \log^2 \frac{s}{M^2}
\right) \nonumber \\ &&
\times \exp \left( - \frac{1}{2} \sum^2_{k=1} \frac{e_f^2}{(4 \pi)^2}
2 \log \frac{s}{\mu M} \log \frac{M^2}{\mu^2} \right) \nonumber \\
&=& {\cal M}^{\rm R}_{\rm Born} \exp \left(- \frac{e_f^2}{(4 \pi)^2c_{\rm w}^2}
\log^2 \frac{s}{M^2} \right) \nonumber \\
&& \times \exp \left( - \frac{e_f^2}{(4 \pi)^2} \left[ \log^2 \frac{s}{\mu^2}
-\log^2 \frac{s}{M^2} \right] \right) \nonumber \\
&=& {\cal M}^{\rm R}_{\rm Born} \exp \left(- \frac{e_f^2}{(4 \pi)^2} \left[
\log^2 \frac{s}{\mu^2} + \frac{s_{\rm w}^2}{c_{\rm w}^2} \log^2 \frac{s}{M^2}
\right] \right) \label{eq:expres}
\end{eqnarray}
where we use the fact that for right handed fermions $T=0, Y= 2Q_f$ and
$g^\prime = \frac{e}{c_{\rm w}}$.
Eq.~(\ref{eq:expres}) thus reproduces the two loop result in Eq.~(\ref{eq:2lres})
where we assume that for small values of the infrared cutoff $\mu \sim \lambda$.

In Ref. \cite{kps}, the same result is obtained for the DL-vertex corrections
to all orders based on a generalization of QCD-results for the electromagnetic
form factor. Also the subleading Sudakov logarithms are in agreement with
Ref. \cite{m} to all orders.

Ref. \cite{cc} is also in agreement with the above results for right handed
particles. It disagrees, however, for left handed particles with the result
obtained from the infrared evolution method.

The only paper which contains a real two loop
calculation in terms of the physical fields
is presented in Ref. \cite{bw}. It uses, other than in the present work, a new
formulation of the Coulomb gauge also for the massive gauge bosons.
The authors of Ref. \cite{bw} find non-exponentiating DL terms at the two loop
level for both right and left handed fermions and trace these terms back to
the fact that there is a mass gap between the neutral gauge bosons
``since $\frac{m_f}{\lambda}\gg \frac{\sqrt{s}}{M} \gg 1$''.

There are several comments in order.
Firstly, our explicit calculation for the right handed case in the Feynman
gauge contradicts their claims, the overall result of course being gauge
independent. Since for right handed fermions the Z-boson
coupling is proportional to $\frac{s_{\rm w}}{c_{\rm w}}$, one can simply
switch off all Z-boson effects by taking the limit $\frac{s_{\rm w}}{c_{\rm w}}
\longrightarrow 0$ as is evident by our result in Eq.~(\ref{eq:2lres}). 
If the result of Ref. \cite{bw} is a consequence of the
aforementioned reasons, the non-exponentiating terms should also 
vanish in that limit (since in the Feynman gauge all
DL-terms from Z-bosons must couple $\sim s_{\rm w}$).
They, however, do not. The authors of Ref. \cite{bw} claim that there are
new uncanceled contributions from so called frog-diagrams\footnote{These
are two-loop two point functions where the external fermions are coupled
to neutral gauge bosons with an induced $W^\pm$-loop in the Coulomb
gauge. In this gauge, all DL corrections are contained in two point
functions unlike the situation for covariant gauges.}(where the
$s_{\rm w}$ terms are indeed canceled by the $W^\pm Z$-vertex)
because of the
massless photon and the massive Z-boson. It should be noted, however, that
those diagrams are infrared finite because the virtualities are cut off
by ${\bf k}_\perp^2 \ge 4 M^2$ in order to produce the $W^\pm$ loop. 
The limit $\lambda \longrightarrow 0$ exists, and
in particular, $\lambda$ does not enter as an argument of a logarithm. 
Thus,
the region producing large logarithms for both neutral gauge bosons is in
fact identical. 
If the mass gap was the origin for the new terms and would vanish if
$\lambda=M$, it
should reveal itself through a term which vanishes in that limit.
While the authors claim that there is exponentiation in that case, their
result shows no sign of a mass gap.
Thirdly, the argument should evidently be independent of the fermion masses.
Thus for massless fermions, i.e. $m_f \sim \lambda$, the abovementioned
reasoning fails since ${\cal O} (1) \sim \frac{m_f}{\lambda} \ll \frac{\sqrt{s}}{M}$ and one expects the effective
theory to be that of the unbroken phase.

In addition the author would also like to point out that for the left handed
case, the authors of Ref. \cite{bw} obtain the same spurious contribution
as in the right handed case plus a term which is non-analytic in the 
coupling $\sim |e_f|$. While non-analytic contributions occur
in bound state problems from restrictions of the effective region of
integration, these terms are disallowed in general scattering amplitudes
by the analyticity properties of S-matrix elements.
The results of Ref. \cite{bw} should thus be rechecked, with a
special focus on the subtleties
of the new Coulomb gauge formulation for massive gauge bosons.

\section{Conclusions} 

In this paper we have calculated the two loop DL corrections to the process
$g \longrightarrow f_{\rm R} {\overline f}_{\rm L}$ using the SM fields. 
It was shown that the
result is in agreement with the calculation based on the infrared evolution
equation method. In particular, the exponentiation found in Refs. \cite{flmm,m}
is confirmed for this process, and it is shown that the mass gap between the
photon and Z-boson does not spoil the higher order Sudakov resummation.
Our result contradicts explicitly recent claims \cite{bw} 
about the breakdown of
exponentiation through mass gap effects. Rather it demonstrates in a 
straightforward way that the matching conditions of the infrared evolution
equation method naturally incorporate the different masses of the neutral
electroweak gauge bosons.

\section*{Acknowledgments}

The author would like to thank E.~Accomando and D.~Graudenz for helpful
discussions and comments.

\end{document}